\documentclass[11pt,a4paper]{scrartcl}

\usepackage{amsthm,amsmath,amssymb}
\usepackage[round]{natbib}% uncomment this for author-year bibliography
\usepackage[colorlinks=true, citecolor=blue, linkcolor=blue, urlcolor=blue]{hyperref}
\usepackage[utf8]{inputenc} %unicode support
\usepackage{mathabx} % for widebar
\usepackage{subfigure}
\usepackage{multirow}

\usepackage{graphicx} % comment if the two lines above are uncommented
\graphicspath{{figs/}} % idem

\newtheorem{theorem}{Theorem}

\newcommand{\esp}{\mathbb{E}}

\begin{document}
\title{Forecasting elections results via the voter model with stubborn nodes}

\author{\textbf{Antoine Vendeville} \\ [2ex]
Department of Computer Science\\Centre for Doctoral Training in Cybersecurity\\
University College London, United Kingdom\\\\
\textbf{Benjamin Guedj} \\ [2ex]
Department of Computer Science\\Centre for Artificial Intelligence \\
University College London, United Kingdom\\
Inria, France\\\\
\textbf{Shi Zhou} \\ [2ex]
Department of Computer Science\\Centre for Artificial Intelligence \\
University College London, United Kingdom\\\\
}
\date{}

\maketitle

\begin{abstract}
We explore a method to influence or even control the diversity of opinions within a polarised social group. We leverage the voter model in which users hold binary opinions and repeatedly update their beliefs based on others they connect with. Stubborn agents who never change their minds (``zealots'') are also disseminated through the network, which is modelled by a connected graph. Building on earlier results, we provide a closed-form expression for the average opinion of the group at equilibrium. This leads us to a strategy to inject zealots into a polarised network in order to shift the average opinion towards any target value. We account for the possible presence of a \emph{backfire effect}, which may lead the group to react negatively and reinforce its level of polarisation in response. Our results are supported by numerical experiments on synthetic data.
\end{abstract}
\section{Introduction}

For decades, modern democratic societies have been polling populations to try and track the popularity of elections candidates and members of governments. Those are often conducted by means of phone, online or even in person surveys, which can be very time-consuming and and usually suffer from limited sample sizes and bias -- \emph{e.g.}\ respondants with controversial views might be reluctant of sharing them. This is why different methods are being investigated nowadays. With the rapid growth of online social platforms such as Facebook or Twitter, any individual can now publicly express their views and opinions, adding to an evergrowing pool of directly accessible data. This has open the door for a new avenue of research, that seeks to use this precious resource to forecast polls and election results without having to survey the population.

As of today, most efforts have focused on applying machine learning methods such as sentiment analysis to evaluate public opinion through samples of Twitter data and try to predict the outcome of democratic processes around the globe \citep{sentiment_aggregate, prednews, spanish2019}. The quality of predictions spans a rather wide range and numerous voices have expressed concerns over these methods, arguing that there are multiple factors at play that may alter their reliability \citep{noyoucannot, digitaltrace2017}. This is why in this work we propose a novel method that does not rely on data analysis but rather uses the authentic and official results of previous elections to perform estimation for future ones. 

More precisely, we consider the well-known voter model for opinion dynamics. A population of connected nodes form a graph where some of them are in state 0 and some others in state 1. Nodes can then randomly change state over time following the distribution of others' states. Nodes are usually meant to represent users on a social network and states their opinions or views. This model then allows to describe in a simple and intuitive manner social dynamics where people are divided between two parties and form their opinion by observing that of others around them. A previous work of ours was dedicated to the theoretical study of this model in the specific case where everyone is influenced by everyone else and some users are stubborn and never change opinion \citep{vendeville2021}. Notably, we provided closed-form expressions for the distribution of opinions at any point in the process and convergence time to equilibrium. 

This paper is a follow-up of that work, as we apply our previous findings to forecast results of both general elections in the United Kingdom and presidential elections in the United States. We consider the share of popular votes won by either the Conservative and Labour party -- or Republican and Democratic party in the US -- as a representation of node states in a fictional graph and perform time-evolving estimation of optimal parameters for the corresponding voter model. This allows us to obtain a theoretical distribution for the number of seats or votes, from which we draw the expected result of future elections. We compare with real-life outcomes to assess the viability of our approach.

\section{Related Literature}

A number of research projects have focused on applying machine learning algorithms to Twitter data in order to forecast opinion poll results or election outcomes. We discuss some of them here and refer the interested reader to \cite{meta_analysis, Phillips2017UsingSM} for more in-depth reviews of the literature. A pioneer work in this area was that of \citet{how140} whose model achieved a mean average error (MAE) of 1.65\% when predicting results of the 2009 German federal election. Authors used Twitter mention counts as an direct indicator of a candidate's popularity, a method that has been considered by several other works as well, often in combination with a sentiment analysis of tweets content \citep{OConnor2010FromTT, sentiment_aggregate, prednews, spanish2019, nigerian2013, multicycle2013, multilevel_italian, towards_passive}. In particular, \citet{prednews} achieved 90\% accuracy in predicting the top two candidates in various municipalities during Brazilian municipal elections, and \citet{sentiment_aggregate} achieved a MAE of 0.63\% when trying to predict opinion poll results during the Portuguese bailout (2011-2014). 

The relevance of such approaches has however been questioned by a number of authors \citep{nigerian2013, multicycle2013, multilevel_italian, towards_passive, pirateparty, digitaltrace2017, noyoucannot}. \citet{pirateparty} showed that merely changing the timeframe of forecast in the work of \citet{how140} would invalidate the results. \citet{nigerian2013} found that use of Twitter mentions mirrored actual popularity of only some of the candidates but not all of them. \citet{digitaltrace2017} argued that mentions count, used in most of the works cited above, show evidence of attention to politics rather that support to the actual candidates. This is why researchers often combine mentions count with sentiment analysis algorithms, but even these can have trouble detecting and correctly interpreting all subtelties of the human language. This particular concern, has been raised by several authors \citep{multicycle2013, multilevel_italian, noyoucannot}. Self-selection, \emph{i.e.} the fact that people choose whether to express their views online or not, may also bias results. Add to it the rife presence of bots on the Twitter platform and it is difficult to say for sure whether the online population is an accurate representation of the real one.

Some researchers have thus considered different avenues, drawing features from the Twitter user graph topology \citep{swedish}, hashtags co-occurences \citep{clinton_trump} or even discarding the social platform entirely and using fluctuations of the Pound to forecast the popularity of the Conservative party in the UK \citep{pound_arima}. Integrating in this line of works, we build a model that does not rely on Twitter but rather uses official results of previous elections to guess the outcome of future ones. Our model is a variant of the celebrated voter model, where nodes on a graph are in one of two possible states and repeatedly update their beliefs to agree with other nodes chosen at random. It was introduced independently by \citet{holley1975} and \citet{clifford_sudbury} in the context of particles interaction. They proved that consensus is reached, \emph{i.e.} that every node is eventually in the same state, on the infinite $\mathbb{Z}^d$ lattice. Several works have since looked at different network topologies: complete graphs \citep{yehuda2002,sood2008,perron2009,yildiz2010}, Erdös-Rényi random graphs \citep{sood2008,yildiz2010}, scale-free random graphs \citep{sood2008,fernley2019}, and other various structures \citep{yildiz2010,sood2008}. Variants where nodes deterministically update to the most common state amongst their neighbours have also been studied \citep{chen2005,mossel2013}.

In this paper we consider the specific case where stubborn nodes who never switch state are present in the graph. Such nodes may for example represent lobbyists, politicians or activists, \emph{i.e.}\ entities looking to lead rather than follow and who will not easily change side. One of those placed within the network can singlehandedly change the outcome of the process \citep{mobilia2003,sood2008}. If several of them are present on both sides, consensus is usually not reachable and instead the distribution of states converge to an equilibrium in which it fluctuate indefinitely \citep{mobilia2007,binary_opinion}. Recently, \citet{mukhopadhyay2020} considered nodes with different degrees of stubbornness and show that time to reach consensus grows linearly with their number. \citet{klamser2017} studied the effect of stubborn nodes on a dynamically evolving graph, and show that the two main factors shaping their influence are their degrees and the dynamical rewiring probabilities. Finally, in our previous work we developed closed-form formulas for the distribution of opinion at any step and convergence time to equilibrium in the case where stubborn nodes are present in a strongly connected network \citep{vendeville2021}. % \citet{mukhopadhyay2020} also show that if one opinion is preferred over the other, consensus is reached on the preferred opinion with a probability that converges to 1 as the network size increases.

\textbf{Our contributions.} In this paper we propose a new model for the forecast of elections outcome, based on official results of previous elections. Our method is based on the voter model with stubborn nodes and uses theoretical results developed in a previous work of ours \citep{vendeville2021}. We apply it to the United Kingdom general elections and in the United States presidential elections and achieve an MAE of 4.74\%. To the best of our knowledge this is the first time such work is conducted. All code used is available online.

%\textbf{Outline.} We lay the theoretical framework in \autoref{theoretical_background}. We apply our findings to the prediction of election results in \autoref{elections_section}. Results for the United Kingdom and the United States are presented in \autoref{uk_results} and \autoref{us_results} respectively. Finally in \autoref{futurework} we conclude and discuss leads for future work. 

\section{Theoretical background} \label{theoretical_background}
%\subsection{The voter model with stubborn agents} \label{voter_model}

Here we present the mathematical framework behind our forecasting method. In the traditional voter model, we consider a group of $n$ nodes labelled $1, \ldots, n$ who are each in state 0 or 1. These states are prone to change over time and we let $x_i(t)$ denote the state of node $i$ at time $t$. Each node has access to the state of some of the others, called its neighbours. Nodes can then be seen as forming a graph of size $n$, with an edge from $j$ from $i$ if and only if $i$ has access to the state of $j$. Here we consider this graph to be a clique with unweighted edges and no self-loops. Thus each node accounts for the state of every other, except their own, with no particular preference. The process then unfolds as follows. Starting with a given initial distribution of states, an independent exponential clock of parameter 1 is associated to each node. Whenever a clock rings, the concerned node changes its state to that of one of its neighbours selected uniformly at random --- or equivalently, chooses its new state by sampling the distribution of its neighbours' states. 
 
We let $N_1(t)$ denote the number of state-1 holders at time $t$; it will be our quantity of interest. Note that the number of state-0 nodes at time $t$ is given by $n-N_1(t)$. We assume $N_1(0)$ is fixed and let $n_1$ denote its value. We are interested in the particular situation where some of the nodes are stubborn, that is never change state, and we describe the evolution of $N_1(t)$ over time. We denote by $s_0$ and $s_1$ the numbers of stubborn state-0 and state-1 nodes respectively and require at least one of them to be strictly positive. To this end we define 
\begin{equation}
	S_n=\{(a,b) \in \{0,\ldots,n\}^2: 0<a+b\leq n\}
\end{equation}
and require $(s_0,s_1)\in S_n$. We write $[m_{ij}]_{i,j}$ to denote the matrix with entry $m_{ij}$ in the $i$-th row and $j$-th column and let $e^M$ denote the exponential of any matrix $M$.

Because $s_0$ and $s_1$ nodes will always be in respective states 0 and 1 no matter what, $N_1(t)$ is comprised between $s_1$ and $n-s_0$ for all $t$. The idea behind our analysis is that it describes a birth-and-death process over the state-space $\{s_1, \ldots, n-s_0\}$ with transition rates, for all $s_1\leq k \leq n-s_0$,
\begin{equation} \label{case3_rates}
	\begin{cases}
		q_{k,k-1} = (k-s_1)(n-k)/(n-1) \\
		q_{k,k+1} = k(n-k-s_0)/(n-1) \\
		q_{k,k} = -q_{k,k-1} -q_{k,k+1}.
	\end{cases}
\end{equation}
Indeed to move from state $k$ to $k-1$ we need a non stubborn state-1 node to adopt the state of an state-0 node. There are $k-s_1$ non stubborn state-1 nodes and for each of these, a proportion $(n-k)/(n-1)$ of the others is in state 0, hence $q_{k,k-1} = (k-s_1)(n-k)/(n-1)$. We obtain $q_{k,k+1}$ via an analogous reasoning and define $q_{k,k}=-q_{k,k+1}-q_{k,k-1}$. Since the process only evolves by unit increments or decrements, $q_{k,j}=0$ if $j \notin \{k-1,k,k+1\}$. As expected we have $q_{s_1,s_1-1}=0$ and $q_{n-s_0,n-s_0+1}=0$. Finally we let $Q=[q_{ij}]_{i,j}$ denote the transition rate matrix. From there we are able to compute the distribution of $N_1(t)$ and its expected value at any point in time.

\begin{theorem} \label{case3_distrib_expect}
	Let $Q$ be the matrix with entries described in (\ref{case3_rates}) and let $N_1(0)=n_1$ be given. Assuming $(s_0,s_1)\in S_n$ is the repartition of stubborn nodes, the probability for $N_1$ to equal $k$ at time $t$ is
	\begin{equation}
	p_{n_1,k}(t) := [e^{tQ}]_{n_1,k}.
	\end{equation}
	Hence,
	\begin{equation} \label{case3_expectation}
	\esp N_1(t) = \sum_{k=s_1}^{n-s_0} k \, p_{n_1,k}(t).
	\end{equation}
	is the expected number of state-1 nodes at time $t$.
\end{theorem}

Because there are stubborn agents in both camps, consensus is never reached and instead the system indefinitely fluctuates within a state of equilibrium. We would like to know if the political system we consider can be considered to be within such state. To this end, the long term expectation of $N_1(t)$ is given by the following theorem.
\begin{theorem} \label{case3_limit}
Assuming $(s_0,s_1)\in S_n$ is the repartition of stubborn agents, the expected number of opinion-1 holders at equilibrium is given by
\begin{equation} \label{esperance_pi}
	\esp \pi = n\frac{s_1}{s_0+s_1}
\end{equation}
where $\pi=(\pi_{s_1}, \ldots, \pi_{n-s_0})$ denotes the steady-state distribution of $N_1(t)$.
\end{theorem}

The theory is developed in another work of ours \citep{vendeville2021} to which we refer the interested reader for more details and proofs.

\section{Setup} \label{methodo}

We use the official database of the United Kingdom general elections results, published by the House of Commons \citep{uk_election}, as well as results for presidential elections in the United States manually collected from Wikipedia.\footnote{\url{https://en.wikipedia.org/wiki/United_States_presidential_election\#Popular_vote_results}} Each time we are interested in the percentage of popular votes won by the two major parties -- Conservative and Labour in the UK, Republicans and Democrats in the US. We assume these quantities correspond to pointwise observations of independent realisations of the voter model. The result of each election can then be forecast via \autoref{case3_distrib_expect}, provided we have an estimate of the quantity of stubborn nodes $(s_0,s_1)$. Thus, our analysis is done in two steps: first we make for each elections an estimate of $(s_0,s_1)$ based on previous results, then \autoref{case3_expectation} gives us the expected value for the coming election that we use as a predictor.

In the UK dataset, our quantity of interest is the percentage of popular votes won by the Conservative and Labour parties in each general elections from 1922 onwards. In the US dataset, it is the number of popular votes gathered by Republicans and Democrats in each presidential elections from 1912 onwards. For the sake of clarity we present our method in the UK case, but note that it directly translates to the US case by replacing Conservative and Labour with Republicans and Democrats.

Because our model cannot account for decimal values values we round the percentages to the nearest integer. Different parties are present, the two major ones being Conservative\footnote{The dataset also includes in Conservative results: National, National Liberal and National Labour candidates for 1931-1935; National and National Liberal candidates for 1945; National Liberal candidates from 1945 to 1970.} and Labour, the rest including Liberal Democrats or Social Nationalists amongst others. Because our model applies to a two-sided situation only, we cannot consider all of them at once. Thus, we aggregate all non-Conservative parties under the label 0 while Conservatives are attributed label 1. We let $x_i$ denote the number of seats won by Conservatives on the $i^{\text{th}}$ elections and $t_i$ the elapsed time, in years, since the starting point 1922. There have been $m=27$ elections total, with the last one taking place in 2019. Thus $t_1=0$ and $t_m=2019-1922=97$. We let $x_m$ denote the percentage of seats won by the conservatives in 2019. To concur with our theoretical framework we consider one seat won by the Conservatives (resp. non-Conservatives) as the observation of an node being in state 1 (resp. 0) amongst $n=100$ of them. The $x_i$'s then correspond to pointwise observations at times $t_i$'s of a realisation of the process $N_1(t)$ described in \autoref{theoretical_background}. All the reasoning described here and in the following will also be applied independently in the cases Labour versus non-Labour, Republican versus non-Republican (US) and Democrat versus non-Democrat (US).

\section{Methodology}
To be able to use \autoref{case3_distrib_expect} to make predictions, we first need to estimate the proportion of potential stubborn nodes in the population, that is the percentage of votes which are guaranteed for or against Conservatives. Let $s_0$ denote the number of stubborn state-0 (non-Conservative) nodes and $s_1$ that of state-1 (Conservative) ones. We look for the values $(s_0^\star,s_1^\star)$ that maximise the log-likelihood of the observed data. Let's say we want to predict results for the $i^{\text{th}}$ election. Because we need at least two datapoints to make an estimaation, we require $3\leq i \leq m+1$. Following the notations introduced in \autoref{theoretical_background} we let $p_{k,l}^{(s_0,s_1)}(t)$ denote the theoretical probability for $N_1(t)$ to go from $k$ to $l$ in $t$ units of time when there are respectively $s_0$ and $s_1$ state-0 and state-1 stubborn nodes. We seek to solve
\begin{align} 
	&\underset{s_0,s_1}{\text{argmax}} \; \sum_{j=1}^{i-2} \log \left( p_{x_j,x_{j+1}}^{(s_0,s_1)}(t_{j+1}-t_j) \right) \label{optim_first}
	\intertext{Indeed, $p_{x_j,x_{j+1}}^{(s_0,s_1)}(t_{j+1}-t_j)$ is by definition the probability for Conservatives to win $x_{j+1}$ percent of the votes in the $(j+1)^{\text{th}}$ election knowing they won $x_j$ percent in the $j^{\text{th}}$ one. Thus we seek to simultaneously maximise the likelihood of all past elections results. Let $Q^{(s_0,s_1)}$ be the matrix with entries calculated via (\ref{case3_rates}). By \autoref{case3_distrib_expect}, we have that (\ref{optim_first}) is equivalent to}
	&\underset{s_0,s_1}{\text{argmax}} \; \sum_{j=1}^{i-2} \log \left[ e^{(t_{j+1}-t_j)Q^{(s_0,s_1)}} \right]_{x_j,x_{j+1}} \label{optim}
\end{align}
The computation of matrix exponential is typically done in cubic time and quickly becomes intractable as the size of the matrix increases. Here however, because we have $n=100$, the number of possible couples $(s_0,s_1)$ is small enough here that (\ref{optim}) can be solved by directly computing the sum for each of these couples individually. The optimal value $s_1^\star$ for $s_1$ then gives us an estimation of the percentage of votes ``locked'' by the Conservative party, proportion of the population that will always root for them. The optimal value $s_0^\star$ for $s_0$ is an estimate of the quantity of such votes for all other parties aggregated.

To make a forecast for the $i^{\text{th}}$ election, we just have to apply \autoref{case3_distrib_expect} with $Q=Q^{(s_0^\star,s_1^\star)}$, $n_1=x_{i-1}$ and $t=t_i-t_{i-1}$. \autoref{case3_expectation} then gives us the expected percentage $\tilde{x}_i$ of votes gathered by Conservatives on that occasion. This can then be compared to the actual value $x_i$ to assess the efficacy of our approach.

\section{Results for the UK} \label{uk_results}

We show in \autoref{stubborn_table} (left) the estimated values for $(s_0^\star,s_1^\star)$, updated with each new election. They seem to globally stabilise between 15 and 25 for both parties. Look at the last value in the Labour case for example, which is $(24,15)$. According to our model, this means there is an estimated proportion of 15\% of voters that will \emph{always} vote Labour. On the other side, 24\% of voters are found to be stubborn ``anti-Labour'' -- by that we don't mean that they are fundamentally against the Labour party but rather that they will \emph{never} vote for it. Note that these estimates fluctuate according to the variability of the data. For example in 1922 and 1923 there were twice in a row 38\% votes for Conservative\footnote{Remember that those value are rounded to the nearest integer to fit the needs of our model -- the actual results were 38.5\% and 38\%.} and as a result it was estimated that 38\% of all voters are stubborn pro-Conservative and the 62\% are stubborn against the party.
This is indeed what maximises the likelihood, with this configuration yielding a probability of 1 for the observed values. On the other hand, with pro-Conservative votes jumping from 38\% to 61\% in 1935, estimated values of $s_0$ and $s_1$ dropped significantly to account for the wide range covered by the data.

\begin{figure}[h!]
	\centering
	\includegraphics[width=.8\textwidth]{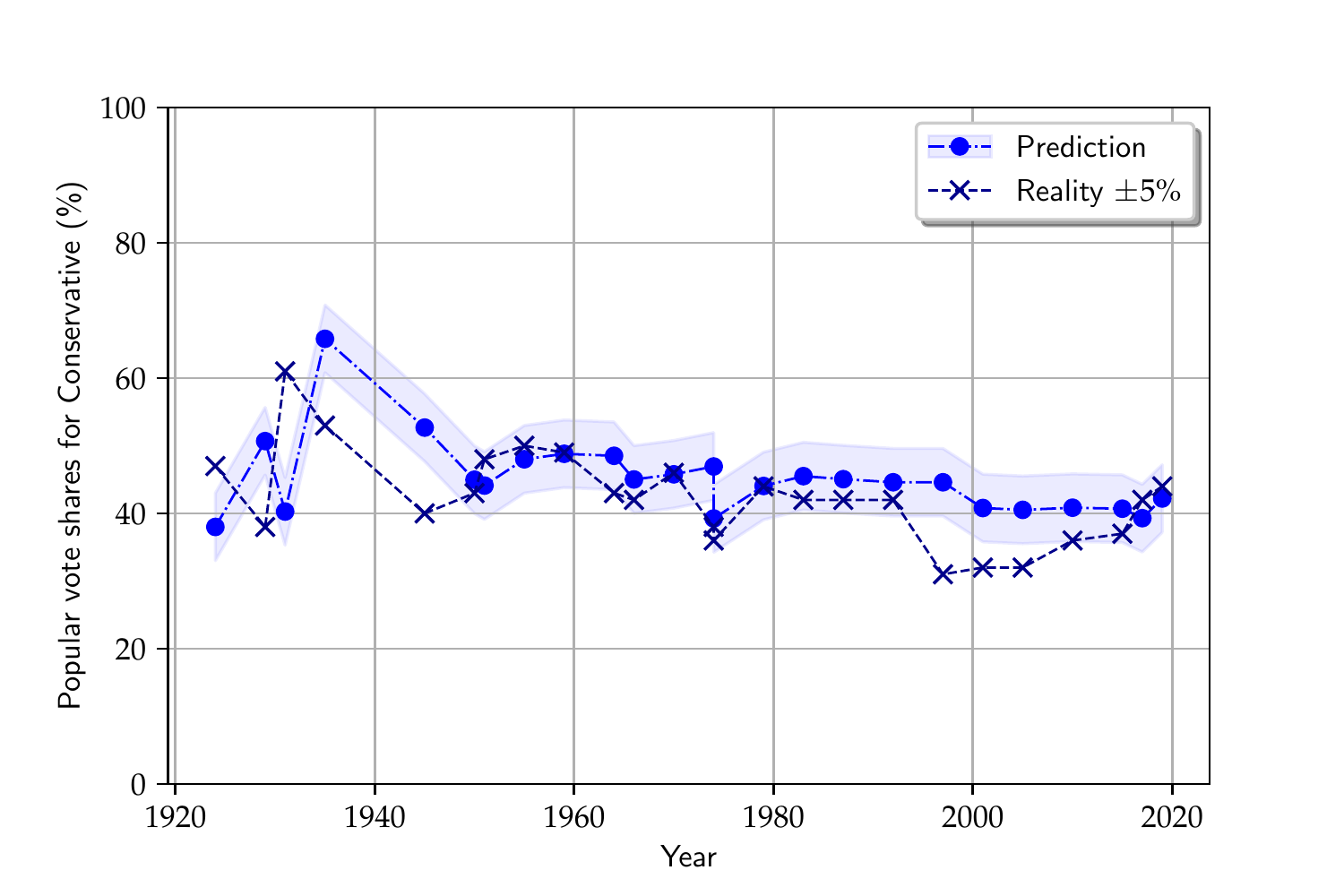}
	\caption{Percentage of popular votes for Conservatives in the UK, prediction and reality. The shaded area covers a $\pm5\%$ deviation away from the predictions.} 
	\label{uk_plot_con}
\end{figure}

\begin{figure}[h!]
	\centering
	\includegraphics[width=.8\textwidth]{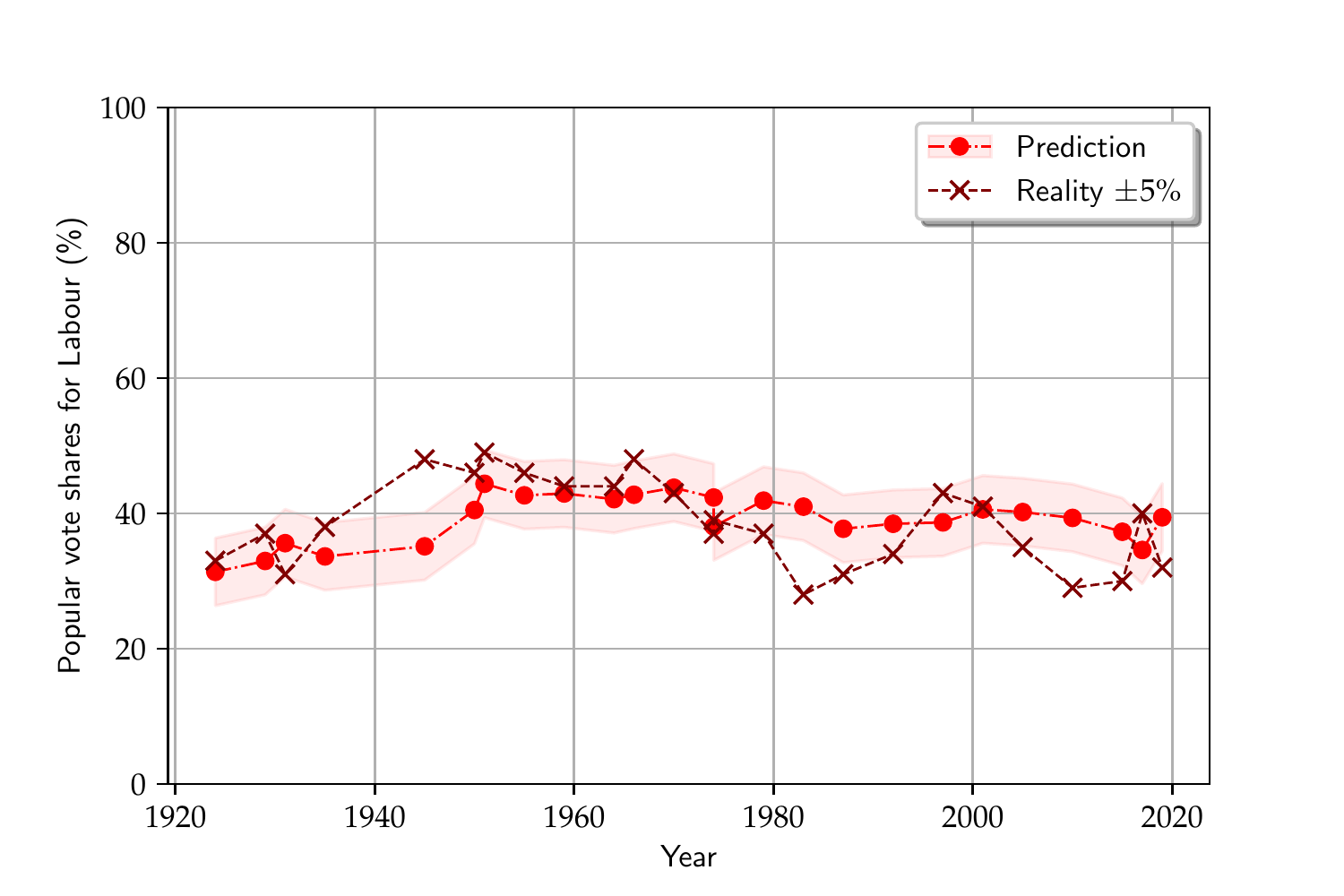}
	\caption{Percentage of popular votes for Labour in the UK, prediction and reality. The shaded area covers a $\pm5\%$ deviation away from the predictions.} 
	\label{uk_plot_lab}
\end{figure}

In \autoref{uk_plot_con} and \autoref{uk_plot_lab} we compare our predictions, that is the expectations $\tilde{x}_i$, with the real outcomes $x_i$. We plot both values for each election starting with the third one that took place in 1924, because the optimisation problem (\ref{optim}) requires $i\geq3$. For both parties, most values seem to fluctuate around the 40\% mark. The global tendency of the real outcomes looks respected by the predictions, albeit with less variability. Also note that most predictions appear to be within a $\pm 5\%$ vicinity of the real values.

\begin{figure}[h!]
	\centering
	\includegraphics[width=.8\textwidth]{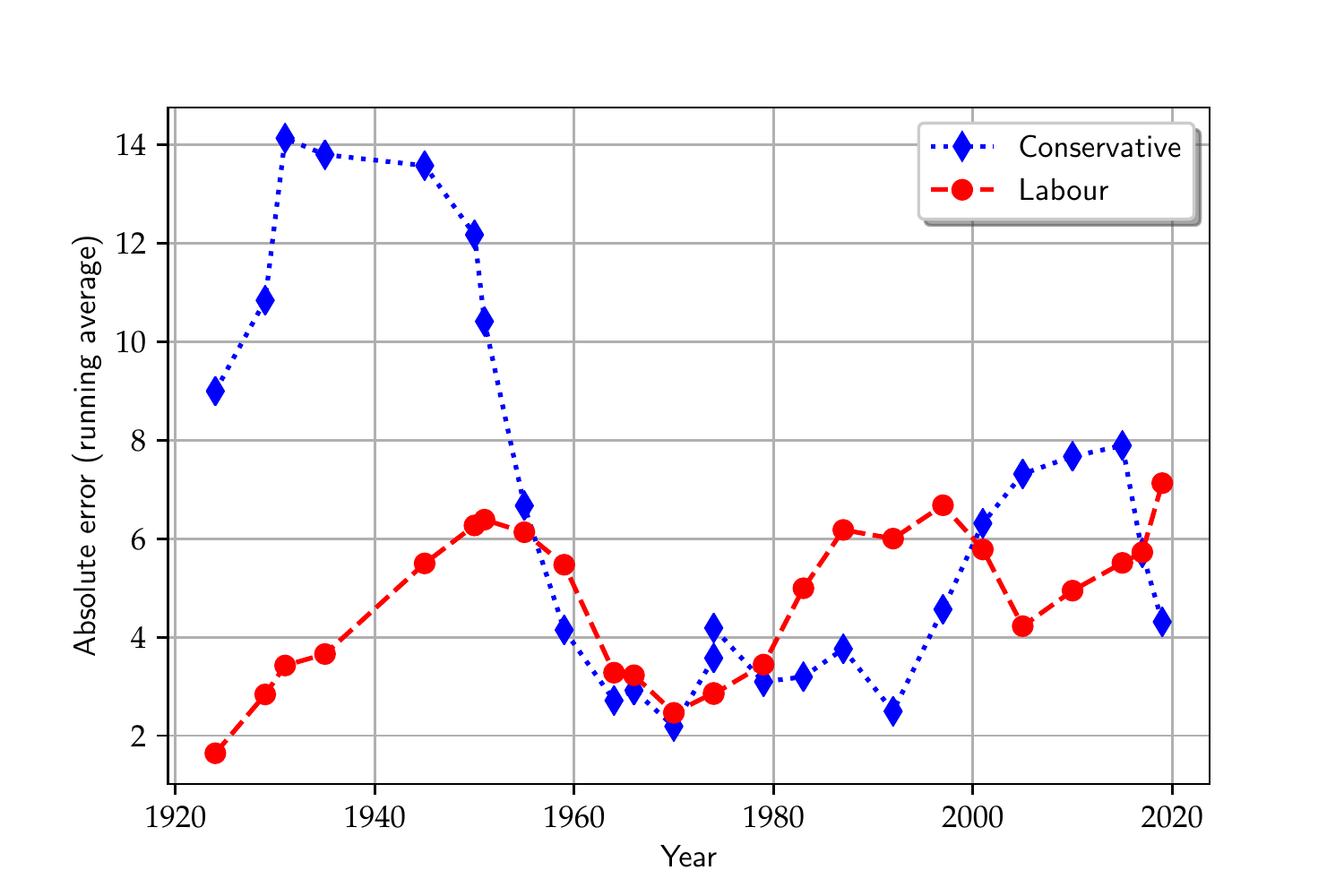}
	\caption{Absolute error between prediction and reality for the UK elections, running average over the last 5 elections.} 
	\label{uk_error}
\end{figure}

To get a better insight we look at the absolute errors $|\tilde{x}_i - x_i|$ of our predictions. We plot running averages over the last 5 elections in \autoref{uk_error}. After a few erratic first years they seem to stabilise between 2 and 8\%. More precisely, if we discard the first few years up until 1960 where the model lacks sufficient amount of data to properly calibrate, we get MAEs of respectively 4.63\% and 5.23\% for Conservative and Labour. Minimal values of 0.06\% for Conservatives in 1979 and 0.40\% for Labour in 2001 are observed, showing that our method was able to make very accurate predictions in these cases. Surprisingly however, the errors do not seem to monotically decrease over time, but rather fluctuate. As a matter of facts, peak absolute errors were observed in 1983 (Labour, 13.0\%) and 1997 (Conservative, 13.6\%).

\section{Results for the US}  \label{us_results} 

We apply the exact method described above to the case of presidential elections in the United States. As we did before we independently consider two cases, Republicans versus non-Republicans and Democrats versus non-Democrats. Presidential elections in the US take place every 4 years and we start with the year 1912, then 1920, 1924, and so on. Here again, keep in mind that due to how the American system work, the party with the most popular votes does not necessarily win the elections. The first estimation we are able to make is based on the first two elections and thus our first prediction is for 1924. 

\begin{figure}[h!]
	\centering
	\includegraphics[width=.8\textwidth]{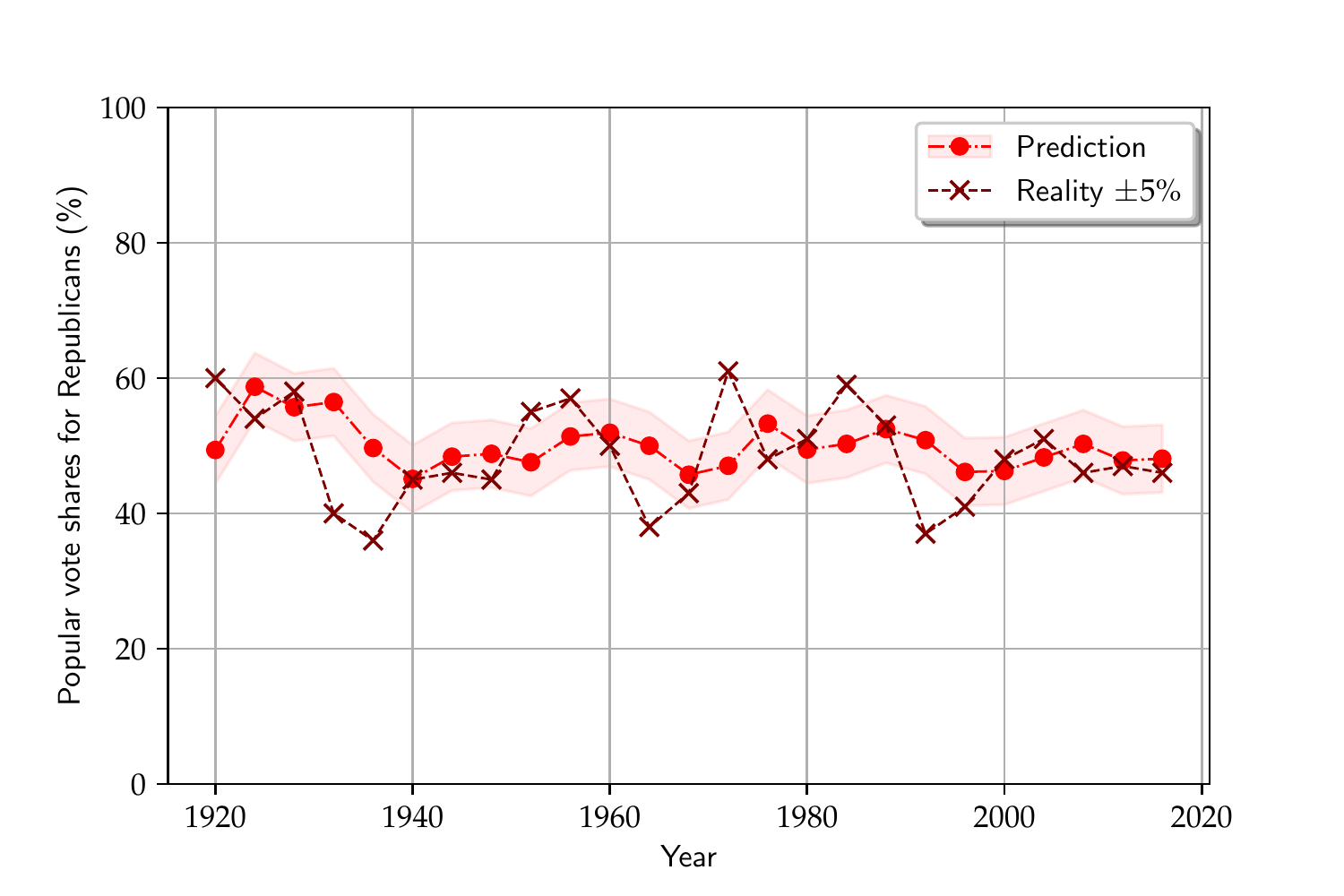}
	\caption{Percentage of popular votes for Republicans in the US, prediction and reality. The shaded area covers a $\pm5\%$ deviation away from the predictions.} 
	\label{us_election_plot_rep}
\end{figure}

\begin{figure}[h!]
	\centering
	\includegraphics[width=.8\textwidth]{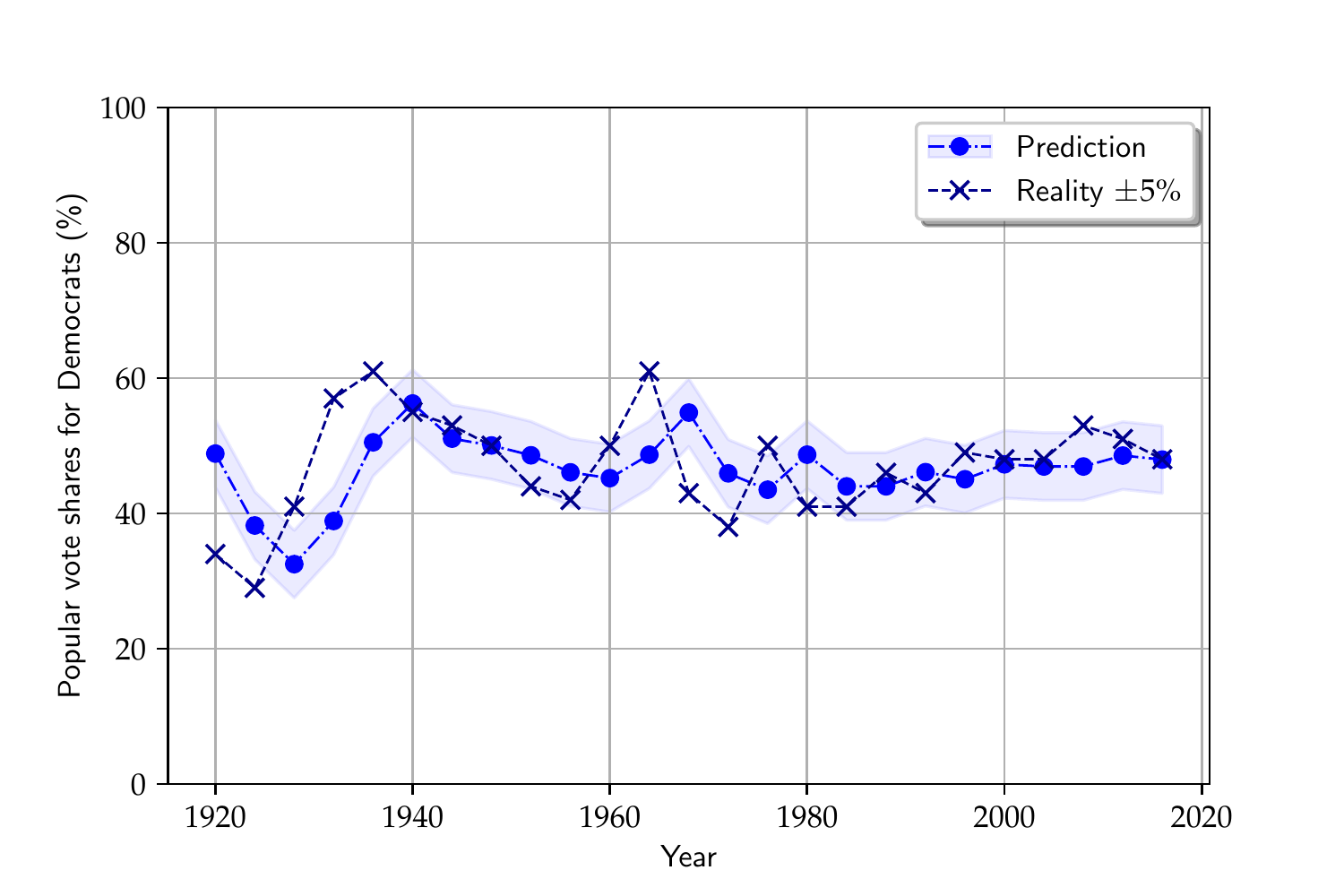}
	\caption{Percentage of popular votes for Democrats in the US, prediction and reality. The shaded area covers a $\pm5\%$ deviation away from the predictions.} 
	\label{us_election_plot_dem}
\end{figure}

We observe similar results as in the UK case. Stubborn values (\autoref{stubborn_table}, right) estimated $(s_0^\star, s_1^\star)$ are close, albeit a little bit lower -- stabilising at (18,17) for Republicans and (16,14) for Democrats. Regarding the predictions (\autoref{us_election_plot_rep}, \autoref{us_election_plot_dem}) we again see a majority of them within a 5\% margin from the actual outcomes, and a prediction curve that looks more stable than the slightly spiky ones with real values. Note that because of the two-party system in place in the United States, both Republicans and Democrats see their share of popular votes fluctuate around the 50\% mark. In the previous case, it was rather around 40\% because of the space occupied by smaller parties such as Liberal Democrats or Scottish National Party amongst others. The two-sided aspect of our model -- always one party (0) versus another (1) -- may thus be more adapted to the study of the US system.

\begin{figure}[h!]
	\centering
	\includegraphics[width=.8\textwidth]{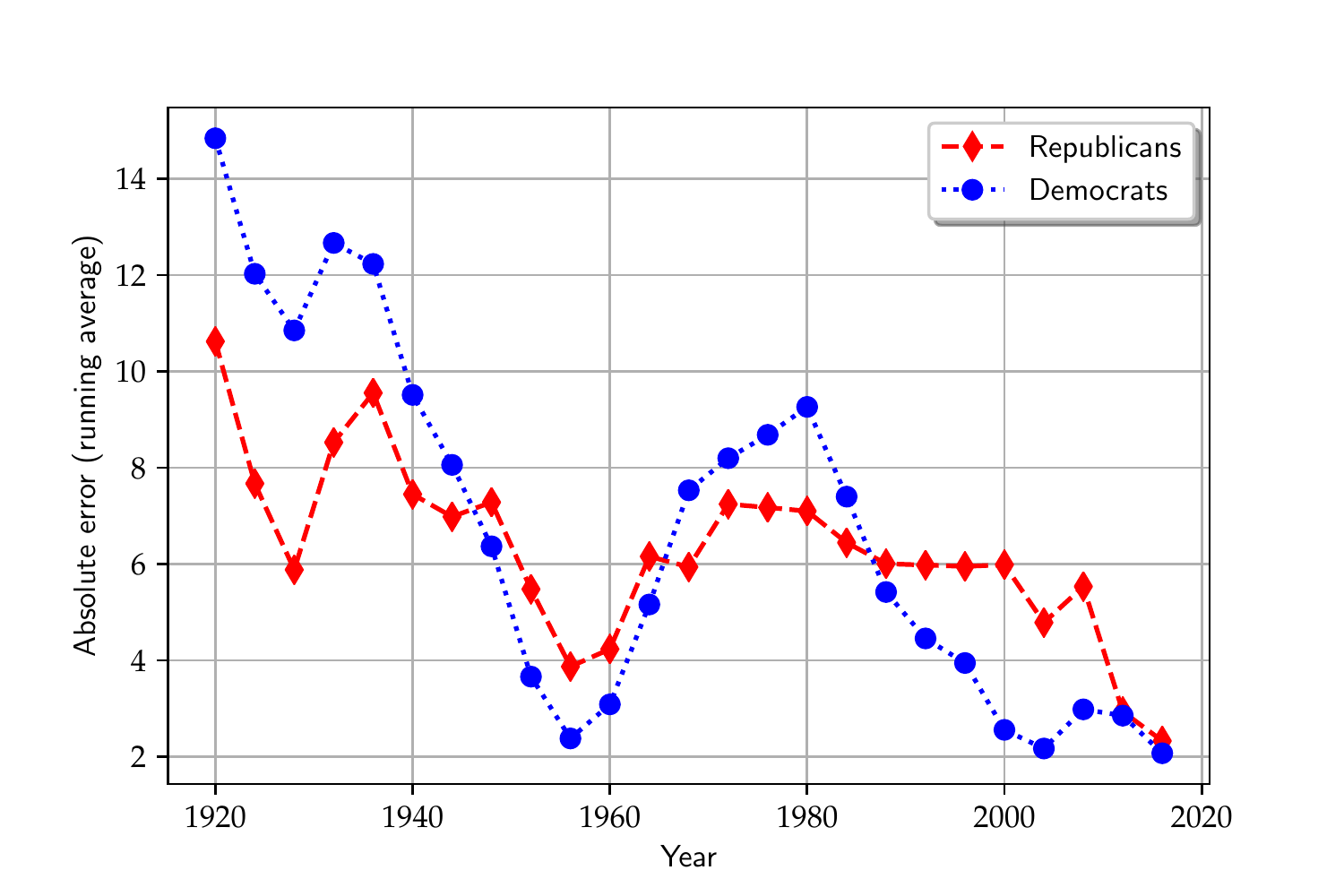}
	\caption{Absolute error between prediction and reality for the US elections, running average over the last 5 elections.} 
	\label{us_error}
\end{figure}

As for the errors, running averages over the last 5 elections are shown in \autoref{us_error}. Here again after a few erratic first years values appear to be comprised between 2 and 8\%. However, where errors in the UK case seemed to increase in the last few years, here they to are dropping down. In fact, our most accurate forecast regarding Democrat votes is for 2016, with only 0.04\% error. For Republicans it is in 1940 with 0.10\%. Peak errors were again around 13\% for both parties, in 1972 (Republicans, 14.0\%) and 1964 (Democrats, 12.3\%). The MAE over all elections, starting in 1940 when forecasts start to stabilise, is 4.27\% for Republicans and 4.83\% for Democrats. This is slightly better than in the UK case (4.63\% and 5.23\%). The MAE error over both cases is then 4.74\%.

%%%% US error table
%\begin{table}[h!]
%\caption{Statistics for absolute error, US elections, rounded to $10^{-2}$.}
%\def\arraystretch{1.25}
%\centering
%\begin{tabular}{l|c|c|c|c|}
%\cline{2-5}
%& min  & max   & mean  & median   \\ \hline
%\multicolumn{1}{|l|}{Republicans} & 0.10       & 16.46        & 5.78         & 4.27     \\ \hline
%\multicolumn{1}{|l|}{Democrats}  & 0.04         & 18.12        & 5.86         & 4.59   \\ \hline
%\end{tabular}
%\label{us_gapstd_table}
%\end{table}

\section{Conclusion and future work} \label{futurework}
In this paper we proposed a new method for the forecast of elections results. A lot of published work have used Twitter data for this purpose, usually applying machine learning algorithm to extract sentiment from tweets and estimate a candidate's popularity this way. Such methods have been criticised in the past few years, with problem ranging from bot presence to text mining reliability that cast doubt over their reliability. As such, our model does not rely on Twitter data at all. Instead, we used official results of past elections in the United Kingdom and in the United States to try and predict outcomes of future ones.

Our method is based on findings from a previous work of ours, where we conducted a theoretical analysis of the voter model with stubborn on strongly-connected graphs. Here we applied those in the case to try and predict the percentage of popular votes won by Conservative and Labour parties in the United Kingdom, and the percentage of popular votes collected by the Republican and Democratic parties in the United States. To do so, we considered official results of past elections as observations of independent realisations of the voter model. From there we were able to perform time-evolving estimates of the model parameters and use them to forecast an outcome. 

Our model yielded an MAE of 4.74\%, reaching absolute errors as low as 0.04\% and as high as 14\%. In their review, \citet{meta_analysis} suggest that any model used to predict the elections outcome should not have an MAE higher than 1 or 2\%. This is because the result of an election is more often than not the matter of just a few percents. According to this standard, our MAE is not low enough to reliably predict the outcome of an election. Some previous works reached error averages as low as 0.63 \cite{sentiment_aggregate} and 1.65\% \cite{how140}. Additionally, we tested our method against the baseline of systematically predict the exact result of the previous election. This simple method returned an overal 5.03\% average error, which is not much worse than the 4.74\% obtained via our method. Moreover, the first few elections results were discarded in both cases, as it was deemed that the model did not have enough data at this point to make predictions with a high enough confidence. The choice of a limit though is made on the basis of our observation of the model's behaviour and is purely subjective. Changing the limit would in turn make for different results that might be better or worse.

Although our method did not prove to yield significant enough results here, we believe it is an interesting step in a novel direction. Despite some impressive accomplishments, there is lot of controversy regarding Twitter-based forecasting which is deemed unreliable by a lot of authors. Thus our method that uses results of previous elections provides a new take on the matter, which only relies on official data. Also our model does not only forecast the elections results, it also gives us estimates of the quantity of stubborn voters that are firmly pro or against any given party. This provides meaningful insight on the political landscape of the considered areas.

Several extensions of the model could be considered to improve its accuracy. First of all, adding in-between election polls to the data would go a long way in improving the estimates. With a few years gap from one election to another, it is too wide a range of possibilites for the model to account for. Second, one could take a deeper look into the past of a country's results adn try to detect tendancies about landslide victories, incumbency reelection and so forth. We believe that having a deeper understanding of the specific country one is working with could substantially improve the model calibration process.

\section*{Funding}%% if any
This project was funded by the UK EPSRC grant EP/S022503/1 that supports the Centre for Doctoral Training in Cybersecurity delivered by UCL's Departments of Computer Science, Security and Crime Science, and Science, Technology, Engineering and Public Policy.

\section*{Abbreviations}%% if any
MAE: Mean Absolute Error. UK: United Kingdom. US: United States.

\section*{Availability of data and materials}%% if any
All code used is available online at \href{https://github.com/AntoineVendeville/HowOpinionsCrystallise}{https://github.com/AntoineVendeville/HowOpinionsCrystallise}. Data for the United Kingdoms elections is available online \citep{uk_election}. Data for the United States elections has been crawled from Wikipedia (\url{https://en.wikipedia.org/wiki/United_States_presidential_election\#Popular_vote_results}).

%\section*{Ethics approval and consent to participate}%% if any
%Text for this section\ldots

\section*{Competing interests}
The authors declare that they have no competing interests.

%\section*{Consent for publication}%% if any
%Text for this section\ldots

\section*{Authors' contributions}
This is a joint work by the three authors, with A.V.\ being the leader of the project.

\section*{Authors' information}%% if any
Antoine Vendeville is a PhD student at University College London (United Kingdom) in the Computer Science department. 

Benjamin Guedj is a Principal Research Fellow in machine learning at University College London and a tenured research scientist at Inria, France. 

Shi Zhou is an Associate Professor at the Department of Computer Science, University College London.

All three authors are affiliated with UCL's Centre for Doctoral Training in Cybersecurity. All three authors are affiliated with UCL's Centre for Artificial Intelligence.

\bibliographystyle{abbrvnat} %unsrt % Style BST file (bmc-mathphys, vancouver, spbasic).
\bibliography{biblio}      % Bibliography file (usually '*.bib' )

\section*{Figures and Tables}

%%% UK plot con

%%% UK plot lab

%%% US plot rep

%%% US plot dem

%%% UK error

%%% US error

\begin{table}[h!]
	\centering
	\caption{Evolution of the estimates for the proportion of stubborn agents $(s_0^\star, s_1^\star)$ over time. Left: United Kingdom. Right: United States.}
	\label{stubborn_table}
	\def\arraystretch{1.25}
	\begin{tabular}{|c|c|c|}
	\hline
	\textbf{Year} & \textbf{Conservative} & \textbf{Labour} \\ \hline
        1924 & (62, 38) & (65, 30) \\ \hline
        1929 & (20, 21) & (61, 30) \\ \hline
        1931 & (28, 20) & (55, 30) \\ \hline
        1935 & (1, 5) & (53, 27) \\ \hline
        1945 & (9, 10) & (48, 26) \\ \hline
        1950 & (11, 10) & (26, 17) \\ \hline
        1951 & (13, 12) & (23, 16) \\ \hline
        1955 & (13, 12) & (23, 16) \\ \hline
        1959 & (15, 14) & (22, 16) \\ \hline
        1964 & (16, 15) & (25, 18) \\ \hline
        1966 & (18, 16) & (25, 18) \\ \hline
        1970 & (18, 16) & (24, 18) \\ \hline
        1974 & (19, 17) & (26, 19) \\ \hline
        1974 & (19, 16) & (26, 19) \\ \hline
        1979 & (19, 16) & (26, 19) \\ \hline
        1983 & (20, 17) & (28, 20) \\ \hline
        1987 & (20, 17) & (22, 15) \\ \hline
        1992 & (22, 18) & (21, 14) \\ \hline
        1997 & (22, 18) & (23, 15) \\ \hline
        2001 & (19, 15) & (24, 16) \\ \hline
        2005 & (18, 14) & (24, 16) \\ \hline
        2010 & (17, 13) & (24, 16) \\ \hline
        2015 & (18, 13) & (22, 14) \\ \hline
        2017 & (18, 13) & (22, 14) \\ \hline
        2019 & (19, 14) & (22, 14) \\ \hline
        2024 & (19, 14) & (24, 15) \\ \hline
        \end{tabular}
        \hspace{30pt}
	\begin{tabular}{|c|c|c|}
	\hline
	\textbf{Year} & \textbf{Republicans} & \textbf{Democrats} \\ \hline
        1920 & (23, 23) & (44, 42) \\ \hline
        1924 & (15, 21) & (18, 12) \\ \hline
        1928 & (18, 23) & (15, 8) \\ \hline
        1932 & (18, 23) & (18, 11) \\ \hline
        1936 & (16, 18) & (10, 8) \\ \hline
        1940 & (13, 13) & (7, 7) \\ \hline
        1944 & (14, 14) & (9, 8) \\ \hline
        1948 & (15, 15) & (9, 8) \\ \hline
        1952 & (17, 16) & (10, 9) \\ \hline
        1956 & (16, 16) & (11, 10) \\ \hline
        1960 & (16, 16) & (11, 10) \\ \hline
        1964 & (17, 17) & (12, 11) \\ \hline
        1968 & (17, 16) & (10, 10) \\ \hline
        1972 & (17, 16) & (12, 11) \\ \hline
        1976 & (15, 15) & (11, 10) \\ \hline
        1980 & (16, 16) & (12, 11) \\ \hline
        1984 & (16, 16) & (13, 11) \\ \hline
        1988 & (16, 16) & (13, 11) \\ \hline
        1992 & (16, 16) & (14, 12) \\ \hline
        1996 & (15, 15) & (14, 12) \\ \hline
        2000 & (16, 15) & (15, 13) \\ \hline
        2004 & (16, 15) & (15, 13) \\ \hline
        2008 & (16, 16) & (15, 13) \\ \hline
        2012 & (17, 16) & (16, 14) \\ \hline
        2016 & (17, 16) & (16, 14) \\ \hline
        2020 & (18, 17) & (16, 14) \\ \hline
	\end{tabular}
\end{table}

\end{document}